\documentclass[12pt]{iopart}

\usepackage{graphicx}
\usepackage{iopams} 
\usepackage{color}

\newcommand{\MeV}{\,\mathrm{MeV}}
\newcommand{\keV}{\,\mathrm{keV}}
\newcommand{\mm}{\,\mathrm{mm}}
\newcommand{\mum}{\,\mathrm{\mu m}}

\newcommand{\Wcm}{\,\mathrm{W/cm^2}}

\newcommand{\ps}{\,\mathrm{ps}}

\newcommand{\MHz}{\,\mathrm{MHz}}
\newcommand{\GHz}{\,\mathrm{GHz}}

\newcommand{\J}{\,\mathrm{J}}

\newcommand{\cc}{\,\mathrm{cm^3}}

\newcommand{\kT}{\,\mathrm{kT}}
\newcommand{\T}{\,\mathrm{T}}
\newcommand{\mT}{\,\mathrm{mT}}

\begin{document}

\title[Laser-driven generation of strong quasi-static magnetic fields]{Laser-driven platform for generation and characterization of strong quasi-static magnetic fields}

\author{J.J. Santos$^1$, M. Bailly-Grandvaux$^1$, L. Giuffrida$^1$\footnote{Present address:
ELI-Beamline Project, Institute of Physics, ASCR, PALS Center, Prague, Czech Republic.}, P. Forestier-Colleoni$^1$, S. Fujioka$^2$, Z. Zhang$^2$, Ph. Korneev$^{1, 3}$, R. Bouillaud$^1$, S. Dorard$^4$, D. Batani$^1$, M. Chevrot$^4$, J.E. Cross$^5$, R. Crowston$^6$, J.-L. Dubois$^7$, J. Gazave$^7$, G. Gregori$^5$, E. d'Humi\`eres$^1$, S. Hulin$^1$, K. Ishihara$^2$, S. Kojima$^2$, E. Loyez$^4$, J.-R. Marqu\`es$^4$, A. Morace$^2$, Ph. Nicola\"i$^1$, O. Peyrusse$^1$, A. Poy\'e$^1$, D. Raffestin$^7$, J. Ribolzi$^7$, M. Roth$^8$, G. Schaumann$^8$, F. Serres$^4$, V.T. Tikhonchuk$^1$,  Ph. Vacar$^4$, N. Woolsey$^6$}

\address{$^1$Univ. Bordeaux, CNRS, CEA, CELIA (Centre Lasers Intenses et Applications), UMR 5107, F-33405 Talence, France}
\address{$^2$Institute of Laser Engineering, Osaka University, Osaka 565-0871, Japan}
\address{$^3$National Research Nuclear University MEPhI, 115409, Moscow, Russian Federation}
\address{$^4$Laboratoire pour l'Utilisation des Lasers Intenses, Ecole Polytechnique, CNRS, CEA, UMR 7605, F-91128 Palaiseau, France}
\address{$^5$Department of Physics, University of Oxford, Parks Road, Oxford OX1 3PU, UK}
\address{$^6$Department of Physics, Heslington, University of York, YO10 5DD, UK}
\address{$^7$CEA/DAM/CESTA, BP 12, F-33405 Le Barp, France}
\address{$^8$Institut f\"ur Kernphysik, Tech. Univ. Darmstadt, Germany}

\ead{santos.joao@celia.u-bordeaux1.fr}

\vspace{10pt}
\begin{indented}
\item[]June 2015
\end{indented}

\begin{abstract}
Quasi-static magnetic-fields up to $800\,$T are generated in the interaction of intense laser pulses ($500\,$J, $1\,$ns, $10^{17}\Wcm$) with capacitor-coil targets of different materials. The reproducible magnetic-field peak and rise-time, consistent with the laser pulse duration, were accurately inferred from measurements with GHz-bandwidth inductor pickup coils (B-dot probes). 
Results from Faraday rotation of polarized optical laser light and deflectometry of energetic proton beams are consistent with the B-dot probe measurements at the early stages of the target charging, up to $t\approx 0.35\,$ns, and then are disturbed by radiation and plasma effects. The field has a dipole-like distribution over a characteristic volume of $1\,$mm$^3$, which is coherent with theoretical expectations. These results demonstrate a very efficient conversion of the laser energy into magnetic fields, thus establishing a robust laser-driven platform for reproducible, well characterized, generation of quasi-static magnetic fields at the kT-level, as well as for magnetization and accurate probing of high-energy-density samples driven by secondary powerful laser or particle beams. Ê
\end{abstract}



\section{Introduction}
The properties of matter on all scales (atoms, molecules, condensed matter, plasmas) are severely modified when exposed to strong magnetic fields (B-fields)\,\cite{Lai_RMP2001}. The possibility of imposing a strong, laser-driven B-field to a variety of samples opens interesting perspectives for laboratory studies of magnetized plasma-\,\cite{Liang_PRL2003}, atomic-\,\cite{Murdin_ncomms2013} and nuclear-physics\,\cite{Lerner_JFusEn2011}. We foresee great progress on the understanding of systems of astrophysical scale\,\cite{Engel_PRA2008, Nordhaus_PNAS2011, Ciardi_PRL2013,  Albertazzi_Science2014, Korneev_PoP2014, Matsumoto_Science2015}, on the improvement of inertial fusion energy schemes\,\cite{Perkins_PoP2013, Strozzi_PoP2012, Wang_PRL2015, Johzaki_NuclFus2015} and for various applications of magnetically-guided particle beams\,\cite{Roth_IPAC2014, Chen_PoP2014}, among many other applications.

State-of-the-art magnets nowadays allow generation of B-fields in the 10-$300\,$T range, depending on their static/pulsed or destructive/non-destructive character\,\cite{Debray_CRPhys2013}. However, reaching, or exceeding the $1\,$kT level, as required for some high-energy-density or atomic physics applications, is much more challenging, unless resorting to large-scale Z-pinch machines\,\cite{Gomez_PRL2014, Schmit_PRL2014} or explosive experiments\,\cite{Bykov_PhysB2001}. Likewise, the heavy technical and infrastructure constraints posed by high-performance pulsed magnets (exceeding 100 T) make them ill-suited to the compactness of laser experiments. This motivates the development of portable, all-optical magnetic generators that may be easily implemented in any high-energy and/or high-power laser facility. 
Relativistic laser interaction with dense targets and the issuing intense currents over the target surface or into the target bulk can generate super-strong B-fields\,\cite{Perez_PRL2013, Korneev_PRE2015}, on the range of $10\kT$ for current short-pulse laser parameters, but these are rather transient as they evolve on the time-scale of $\sim 10\,$ps. 
Quasi-static magnetic-field production coupled to laser facilities has been explored so far by the development of capacitor-bank pulsed discharges in solenoids (magnetic pulsers), but the specific physical limitations restrain the maximum generated fields to the range of $\sim 40\,$T (lowered to a more safe level$\sim 20\,$T in usual operation)\,\cite{Albertazzi_Science2014, Albertazzi_RSI2013}.    

Instead, the use of powerful lasers interacting with so-called capacitor-coil targets - first proposed by Daido\,\etal \,back in the 1980s\,\cite{Daido_PRL1986, Courtois_JAP2005} and recently explored to higher levels by Fujioka\,\etal\cite{Fujioka_SciRep2013} - give unprecedented quasi-static (time-scale of a few ns) high B-field amplitudes for such a compact system ($\sim\,$mm) and an energy laser pulse driver of $1\,$kJ: they reported a B-field of $\approx 1.5\,$kT at $0.65\,$mm away from the U-turn coil centre, measured by Faraday rotation of the polarization of a probe laser beam in a $\mathrm{SiO}_2$ sample. But the reported value would yield, according to a magnetostatic simulation of the U-shaped capacitor-coil target \cite{Radia}, a non realistic magnetic energy, greater than the invested laser energy. The problem probably lies in that the tabulated Verdet constants of the birefringent crystals are of questionable validity in the presence of strong and rapidly changing B-fields, or maybe the crystal properties begin to be affected by X-rays and fast particles due to the close laser-target interaction. 

Besides reaching high B-field strengths, advantages of laser-driven coil-targets is that they can be of relatively low cost if mass produced and are adaptable to laser sources with higher repetition rate. Most importantly these coils have an open geometry providing easy access for several diagnostic views and to the magnetization of secondary samples eventually driven by secondary laser or particle beams.  
In this study, we present an accurate characterization of the B-field produced by laser-driven capacitor-coil targets, using i) high-frequency pickup B-dot probes, ii) Faraday rotation of laser probe beam polarization (we placed the birefringent crystals sufficiently away from the coil to measure B-fields weaker than those reported by Fujioka\,\etal\cite{Fujioka_SciRep2013}, but using much more sensitive crystals) and iii) deflectometry of an energetic proton beam. This combination of three independent diagnostics confirms that high power laser facilities may be used for B-field production in the range of several hundred Tesla with a controlled and reproducible rise time, peak value and spatial distribution.

\section{Methods and experimental results}

\begin{figure}
\center
\includegraphics[width=150mm]{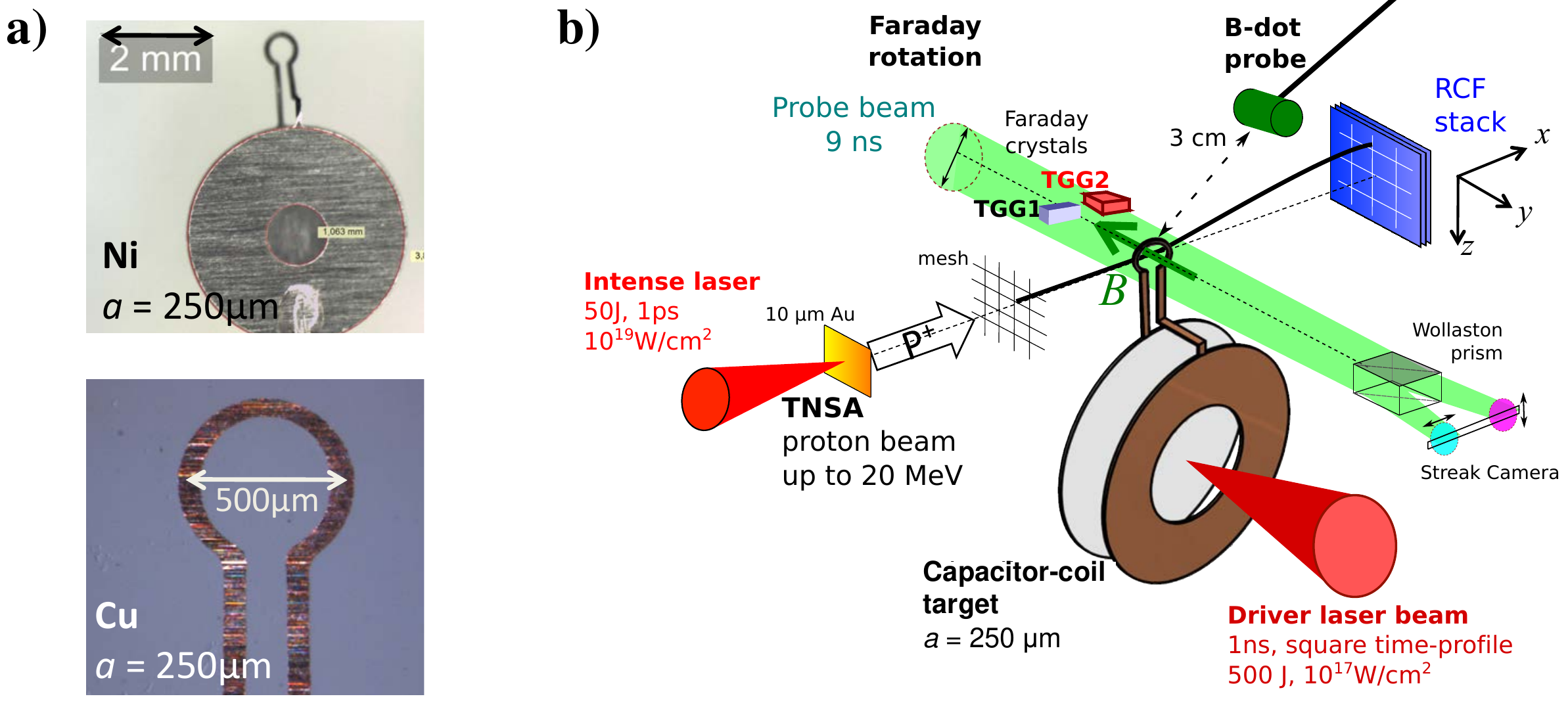}
\caption{\small (color online) {\bf a)} Photographic views of the capacitor-coil targets. {\bf b)} Sketch of the experimental setup. \label{Fig-setup}}
\end{figure}

The experiments were conducted at the LULI pico 2000 laser facility with a $1.057\mum$ wavelength ($1\omega_0$), $500\pm30\,$J, $1\,$ns flat-top long-pulse laser beam ($\approx 100\,$ps rise time), focused to intensities of $(1.0\pm0.1)\times10^{17}\Wcm$. The targets were made of two parallel disks ($3500\mum$ diameter, $50\mum$ thickness, with a hole in the front one enabling focusing of the laser pulse into the rear disk's surface), connected by a coil-shaped wire (coil radius $a=250\mum$, squared rod section of $50\mum\times50\mum$): see Fig.\,\ref{Fig-setup}. The targets were made of Cu, Ni or Al ($1^{\mathrm{st}}$ experiment, with a $1750\mum$-diameter hole in the front disk), or exclusively of Ni ($2^{\mathrm{nd}}$ experiment, with a $1000\mum$-diameter hole). The target parameters were reproducible within a $\pm1\mum$ precision thanks to accurate laser cutting of an initial 2D metallic form, the only effectively variable parameter being the distance between the disks, $d_0 = 900 \pm 200\mum$, as a consequence of the manual target folding. 

The laser pulse irradiates the rear disk passing through the hole of the front disk and creates the supra-thermal electrons that are escaping the potential barrier\,\cite{Dubois_PRE2014, Poye_PRE2015}. A fraction of them are captured by the opposite disk. The target reacts like an RL-circuit to the potential difference between the disks and the subsequent discharge current through the coil-shaped wire. The laser-driven target charging and the discharge current process simultaneously (steady-state regime), during the laser-pulse irradiation, establishing a quasi-static current $I$ looping in the space between the disks and the connecting wire. The coil, of radius $a$, concentrates the magnetic flux yielding a quasi-static, dipole-like B-field over a time-scale of a few ns. The amplitude of the B-field near the coil centre scales like $B_0 \approx \mu_0I/2a$, where $\mu_0$ is the vacuum permeability. The target geometry, namely the distance between the disks, defines the short-circuiting time of the system, which is of the order of 1 - $2\,$ns: It happens when the plasma plume ejected from the irradiated disk reaches the opposite disk or when thermally expanding conducting wires start to overlap. This time is likely to be reduced by the front disk thermal expansion induced by X-ray irradiation from the rear disk plasma. 

The experimental setup is sketched in Fig.\,\ref{Fig-setup}-b). 

\subsection{B-dot probing}

The B-dot probe axis was positioned parallel to the target's coil axis, either in the plane of the coil at a $30\pm0.5\mm$ distance from the coil centre and at a $33^{\circ}$ angle above equator ($1^{\mathrm{st}}$ experiment), or approximately along the coil axis at $70\pm1\mm$ distance from the coil centre, with a $10\pm0.5\mm$ horizontal offset to clear the path for the probe beam used for Faraday rotation and shadowgraphy ($2^{\mathrm{nd}}$ experiment). 
The B-dot probe (and associated electronics) has a $2.5\,$GHz acquisition bandwidth, and the chosen sampling frequency yielded signals with time resolution of $\approx 50$ and $\approx 10\,$ps respectively in the $1^{\mathrm{st}}$ and $2^{\mathrm{nd}}$ experiments. 

\begin{figure}
\center
\includegraphics[width=150mm]{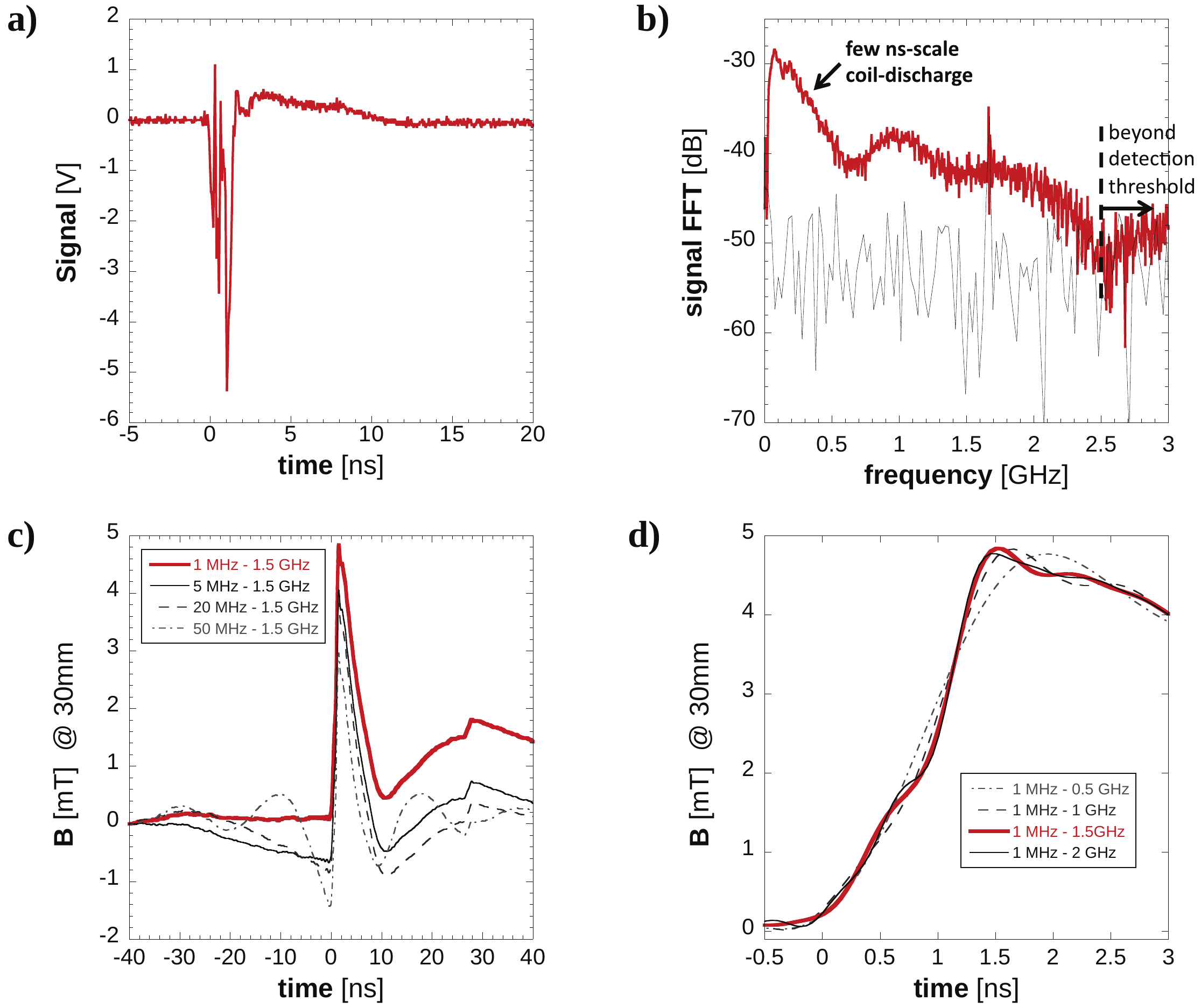}
\caption{\small (color online) Sample B-dot signal and analysis obtained on the 1st experiment, for a Cu-$a=250\mum$ target, with the probe at $30\mm$ from the coil centre. {\bf a)} Signal measured by the $2.5\GHz$ bandwidth detection system. {\bf b)} FFT of the signal (thicker red curve). The thin black curve is the spectrum noise evaluated from the signal FFT for $t<-50\ps$. {\bf c)}, {\bf d)} B-field signal after integration of the B-dot signal: thicker red curves for the default frequency bandpass from $\nu_{\mathrm{min}}=1\MHz$ to $\nu_{\mathrm{max}}=1.5\GHz$, thinner black curves for test variations on {\bf c)} $\nu_{\mathrm{min}}$ and {\bf d)} $\nu_{\mathrm{max}}$. \label{Fig-BdotSample}}
\end{figure} 

Figure \ref{Fig-BdotSample} details the analysis of a signal obtained in the $1^{\mathrm{st}}$ experiment with a Cu target: a) A raw detected signal (attenuators excluded), proportional to the time-derivative of the B-field at the probe position, b) the signal corresponding spectrum, calculated by the Fast-Fourier-Transform (FFT) of the signal (thick red curve; the thin black curve is the FFT of the signal for $t<-50\ps$, that is the noise spectrum). 

\begin{figure}
\center
\includegraphics[width=150mm]{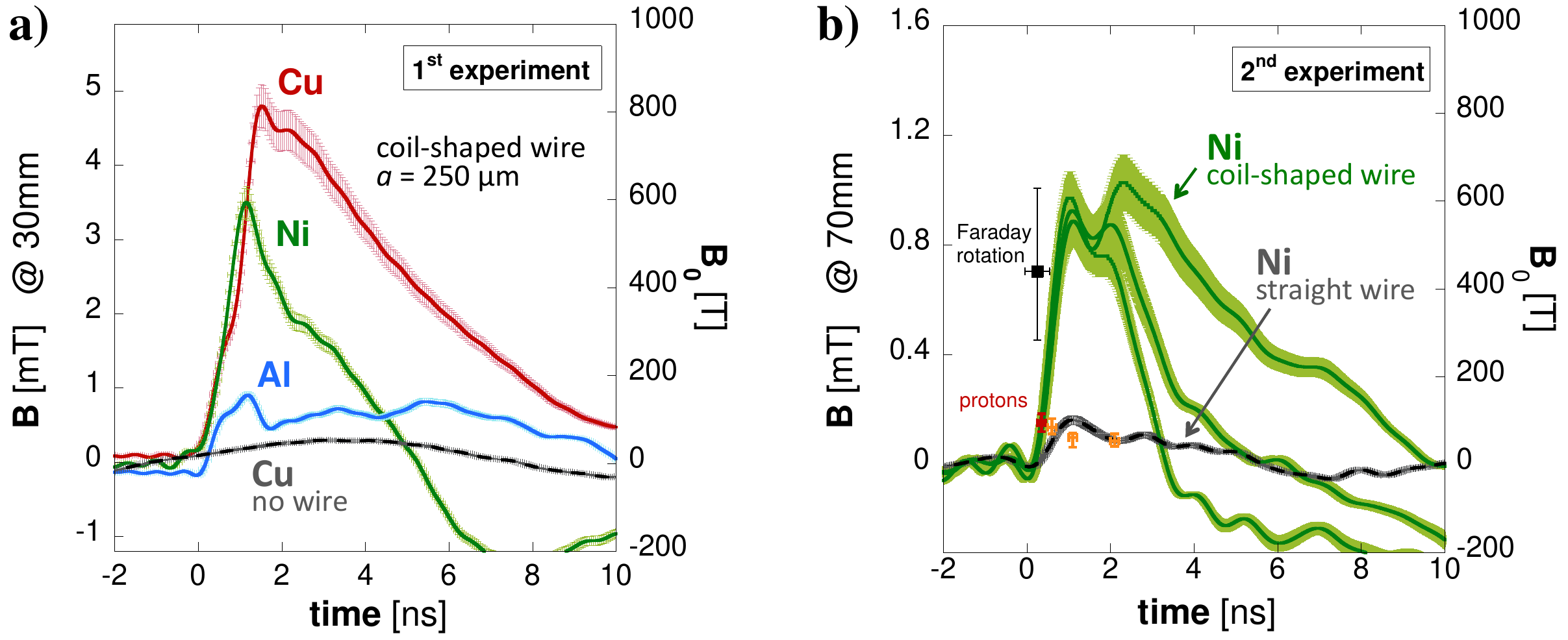}
\caption{\small (color online) Capacitor-coil target results of B-field against time measured by the B-dot probes (solid curves), for the B-field at the probe positions (left-hand-side ordinates) and the corresponding $B_0$ values at the coil's centre (right-hand-side ordinates): {\bf a)} $1^{\mathrm{st}}$ experiment - targets of different materials. {\bf b)} $2^{\mathrm{nd}}$ experiment - Ni targets. The grey dashed curves correspond respectively to a shot on the rear disk holding the two Cu disks parallel at the distance $d_0=900\mum$ but without any connecting wire, and to a Ni target with a straight wire between the disks (no coil): their values refer exclusively to the B-field at the probe position (left-hand-side ordinates). The symbols refer exclusively to the B-field at the coil centre, $B_0$ (right-hand-side ordinates), and are the measurements obtained for Ni capacitor-coil targets by Faraday rotation (square, at $t=0.2\,$ns) and by proton-deflectometry (red circle, at $t=0.35\,$ns). The smaller orange circles represent B-field estimates from proton-deflectometry images obtained at later times: the discrepancy with B-dot probe results is explained in the text by electrostatic effects due to electron trapping near the coil. \label{Fig-allData}}
\end{figure}

Figures \ref{Fig-BdotSample}-c) and d) show the temporal evolution of the B-field, on the level of a few mT. This was obtained by integration of the signal in panel a) for different configurations of bandpass filtering: the optimized result, represented by the thicker red curves, was obtained with the minimum frequency of $\nu_{\mathrm{min}}=1\,$MHz to cut the DC component and $1/\nu$ noise, and a maximum frequency of $\nu_{\mathrm{max}}=1.5\,$GHz to cut the high frequency parasitic EMP emission from higher frequency ground discharge currents\,\cite{Dubois_PRE2014, Poye_PRE2015} ($1^{\mathrm{st}}$ experiment;  we used $\nu_{\mathrm{max}}=1\,$GHz for the $2^{\mathrm{nd}}$ experiment data, according to the reached spectral resolution). Thinner curves correspond to variations either on $\nu_{\mathrm{min}}$ [panel c)] or on $\nu_{\mathrm{max}}$ [panel d)] testing the result sensibility to these parameters: Raising of $\nu_{\mathrm{min}}$ yields an offset of the B-field at $t=0$ (start of laser irradiation), yet the amplitude of the signal is consistent for $\nu_{\mathrm{min}}$ up to $20\,$MHz. For fixed $\nu_{\mathrm{min}}=1\,$MHz, the B-field rise-time and maximum yield do not significantly change for $\nu_{\mathrm{max}}$ down to $1\,$GHz. Therefore, the large spectral peak at a few hundreds MHz in panel b) corresponds to the main B-field signal due to the target discharge through the coil-shaped wire, the second large peak around $1\,$GHz determining the second B-field peak oscillation at $t\gtrsim 2\,$ns (see other measurements in Fig.\,\ref{Fig-allData}).   

\begin{figure}
\center
\includegraphics[width=150mm]{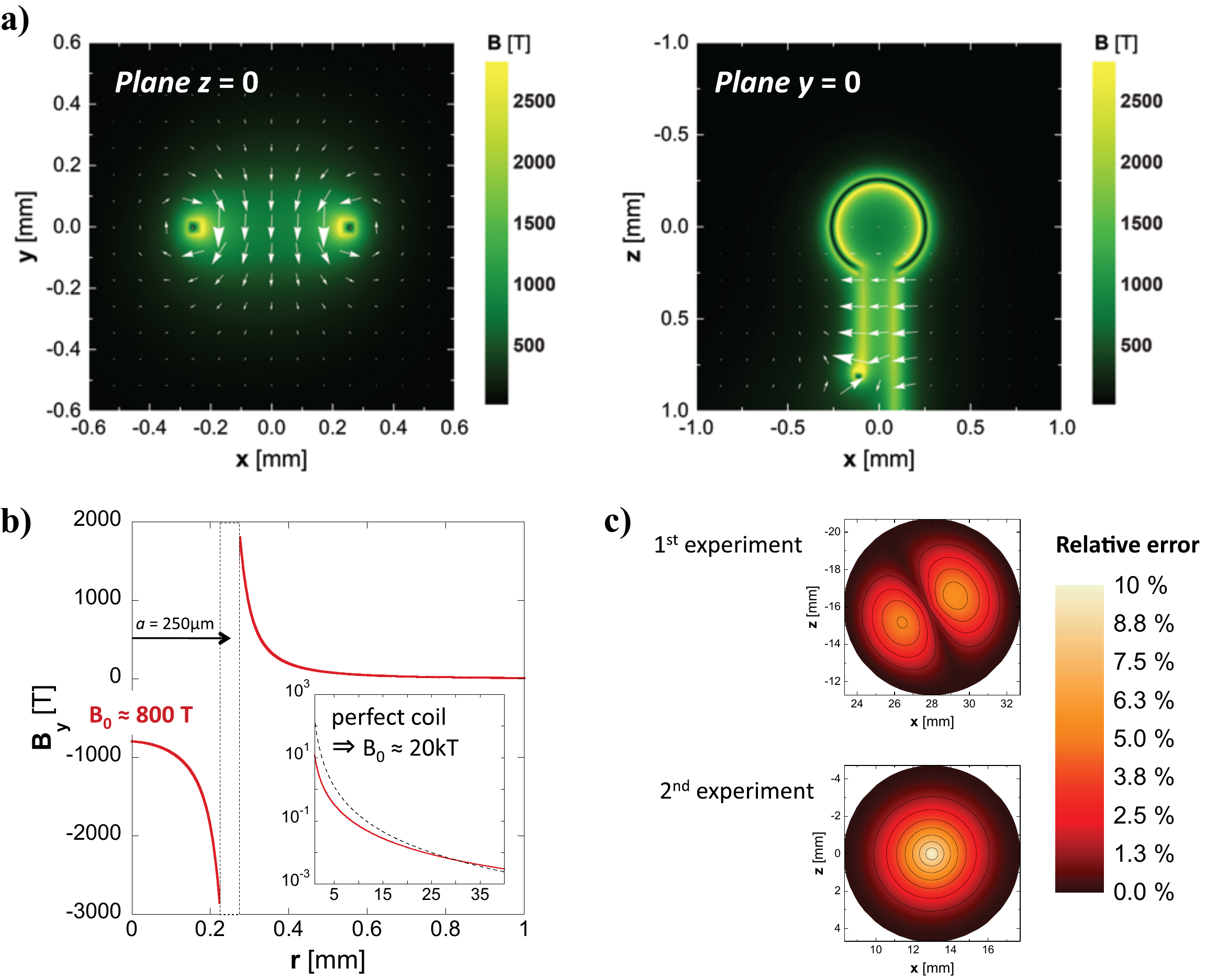}
\caption{\small (color online) {\bf a)} 2D projections of the 3D Radia magnetostatic code extrapolation of the B-field produced by a $a=250\mum$ capacitor-coil target for a circulating current $I=340\,$kA: plane $z=0$ on the left, plane $y=0$ on the right (the coil centre is at the frame origin). {\bf b)} Corresponding axial component of the B-field against distance from the coil centre, along the line between coil centre and the B-dot probe position for the $1^{\mathrm{st}}$ experiment, for a capacitor-coil target (solid red curve) and for a perfect coil of the same radius $a$ (dashed black curve), for the same current. {\bf c)} Relative incertitude of the B-field measurements by the B-dot probes at their positions, respectively in the $1^{\mathrm{st}}$ (top) and the $2^{\mathrm{nd}}$ (bottom) experiments. \label{Fig-B-fieldmap}}
\end{figure} 

The 3D magnetostatic code Radia \cite{Radia} was used to simulate the spatial distribution of the B-field taking into account the target and coil geometry, and their connection wires. 
We used the current $I$ as a free parameter, varying it to obtain an equality between the measured and simulated B-field values at the B-dot spatial positions. The boundary condition consisted in imposing a closed circuit with the same current in the coil and between the two disks. Figure \ref{Fig-B-fieldmap}-a) represents two perpendicular planar sections of the B-field map, one horizontal containing the coil axis, $z=0$ (left), the other vertical corresponding to the coil plane, $y=0$ (right) [see Fig.\,\ref{Fig-setup}-b) for the axis orientation: the $y$-axis is the coil axis and the origin is here at the coil centre]. The field was calculated with an injected current of $I=340\,$kA and a coil radius $a=250\mum$, yielding the measured B-field peak value for the Cu target (given the $\approx 10\,\mathrm{\mu m/ns}$ coil rod expansion velocity measured by time-resolved optical shadowgraphy, no rod expansion was considered in the magnetostatic calculations): the B-field norm is on color scale, the arrows represent the B-field vector projections on the two planes. The spatial distribution on the $z=0$-plane clearly evidences, as expected, a dipole-like B-field. One can also see on the $y=0$ plane that the B-field norm is quite homogenous over the space inside the coil. Poloidal fields around the coil rod and straight parts of the wire are quite strong, and they mostly determine the B-field distribution below the coil region.  

For the same calculation, Fig.\,\ref{Fig-B-fieldmap}-b) shows the amplitude of the B-field component parallel to the coil-axis as a function of the perpendicular distance, along the line connecting the coil and the B-dot probe centres, respectively at $r=0$ and $r=30\mm$ (solid red curve). The vertical dashed lines delimit the $50\mum$-thick coil rod position.
For a measured $5\mT$ at $r=30\mm$, the extrapolated value at the coil centre is $B_0=800\,$T for the capacitor-coil target, with the uncertainty range of $[795, 825\,\mathrm{T}]$ accounting for target-to-target eventual variations of $\pm 2\mum$ on the separation between the two connection points of the wire circular part with the vertical rods. An over-estimated extrapolation B-field value of $B_0=20\kT$ (for a $I=8.5\,$MA current) is obtained when simulating a perfect circular coil of the same radius (dashed black curve in the insert): such unrealistic value illustrates the importance of an accurate modeling of the real target, in particular the discontinuity on the coil circular part. Yet, our target production and characterization are sufficiently precise to have only a very small contribution to the result uncertainty.  

As for the B-dot probes distance with respect to the coil, Fig.\,\ref{Fig-B-fieldmap}-c) shows the calculated absolute relative uncertainties of the B-field measurements at the probe positions: the values are plotted over the probes circular section and account for the 3D B-field gradients over the probes cylindrical volume. The estimated B-field errors, which remain under the $6\%$ and the $10\%$ respectively for the $1^{\mathrm{st}}$ and $2^{\mathrm{nd}}$ experiments (two different probe orientations and distances relative to the coil, as described above), are mainly determined by the gradients along $x$ and $z$ in the first case (top), and along $y$ for the second case (bottom), as expected from the setup geometry.   

The typical final results for the produced B-field as a function of time are summarized in Fig.~\ref{Fig-allData}. The solid curves are the B-dot probe results obtained with capacitor-coil targets for the B-field at the probe positions (left-hand-side ordinates) and the corresponding $B_0$ values at the coil's centre (right-hand-side ordinates, as extrapolated from the Radia magnetostatic calculations).
In the $1^{\mathrm{st}}$ experiment, panel a), the peak values of the B-field depend on the target material, yielding $B_0\approx 800$, $\approx 600$ and $\approx 150\T$ ($\pm 6\%$), respectively for Cu (red), Ni (green) and Al targets (cyan).
In the $2^{\mathrm{nd}}$ experiment, panel b), the B-dot probe measurements are showing a shot-to-shot reproducible charging using Ni targets: The B-field production is synchronous with the beginning of the laser irradiation (at $t=0$, within a $\pm 100\ps$ calibration incertitude) and the rise time of $\approx 1\,$ns is consistent with the duration of the laser pulse (this is the system charge time). The reproducible measurements of a peak B-field of $\approx 1\,$mT are extrapolated to the coil centre peaking at $B_0\approx 600\T$ ($\pm 10\%$). This is in very good agreement with the result obtained with the Ni target in the $1^{\mathrm{st}}$ experiment, where the probe was at a different position relative to the coil. 

The targets used in the $2^{\mathrm{nd}}$ experiment had a reduced front disk hole (diameter of 1000 instead of $1750\mum$), seeking for capturing more non-thermal electrons. However, a comparison of the results leads to the conclusion that the hole size is not a determining parameter for the peak B-field strength, yet the second B-field peak oscillation at $t\gtrsim2\,$ns is more pronounced. 
The typical duration of the pulsed B-field is of a few ns, with fluctuations mostly due to the varying distance $d_0$. All signals present slower amplitude variations at later times (non represented), probably corresponding to electromagnetic coupling with objects around the target (secondary targets, diagnostics, chamber walls), converging to zero at $\sim150\,$ns. 
 
The efficiency of the coils was tested by B-dot measurements in two {\it blank} shots: grey dashed curves in Fig.~\ref{Fig-allData}-a) and b), corresponding respectively to two Cu disks without any connecting wire (held separately), and to a Ni target with a straight wire between the disks (no coil). The dashed curves' values correspond exclusively to the measurements at the probe positions (left-hand-side ordinates). One concludes that i) the laser-target interaction and the escaping high-energy electron currents have a negligible contribution to the measured signals, and ii) the contributions from currents flowing through other target segments besides the coil (in particular the straight parts of the wire) are sufficiently weak if compared to the results obtained with coil-shaped wires: the coil is the dominant source of the magnetic flux. 

The B-dot results for $t\lesssim 0.4\,$ns were confirmed in the $2^{\mathrm{nd}}$ experiment by measurements of the Faraday rotation of the polarization direction of a linearly-polarized probe laser beam, and of the laser-accelerated proton deflections: respectively full square and circular-symbols in Fig.\,\ref{Fig-allData}-b). The B-field peak values could not be measured by these two diagnostics due to laser-plasma effects, as described in the next sections.   

\subsection{Faraday rotation}

The Faraday rotation measurements, using a $9\,$ns-duration probe laser at $533\,$nm wavelength incident along the coil-axis, were performed with two $500\mum$-thick birefringent Terbium Gallium Garnet (TGG) crystals with its centre placed at $3.5\mm$ from the coil plane: TGG1 ($0.5\,$mm-wide) was centred on the coil axis and TGG2 ($1\,$mm-wide) at a $1.9\,$mm perpendicular offset. The TGG Verdet constant, $11.35^{\circ}\mathrm{/T/mm}$, is 38 times higher than that of $\mathrm{SiO}_2$ used in the earlier work by Fujioka\,\etal\,\cite{Fujioka_SciRep2013}, allowing to be sensitive to weaker B-field strengths. No Faraday rotation measurements were successful with the crystals located neither closer to the coil, because of signal blackout quasi-synchronous with the laser irradiation of the target - due to very rapid crystal ionization by the hard X-rays and fast particles issuing from the interaction region - nor further away from the coil where the local B-field was too weak. 
The Faraday effect was measured by using a time-resolved polarimeter, constituted of a Wollaston prism to separate the two perpendicular components of the probe beam field, and a streak camera [see Fig.\,\ref{Fig-setup}-b)]. The time-resolution of $\approx 300\ps$ was limited by the camera time-resolution and by the laser jitter. The quantification of the rotation angle was previously calibrated, without B-field, by quantifying the polarization ratio of the two transmitted perpendicular-polarization signals, defined as $R = E_{\parallel}^2  / (E_{\parallel}^2+E_{\perp}^2)$, as a function of the orientation of the incident light polarization. The obtained linear fit of the experimental data was consistent with the Malus law by an average $\chi^2\approx 0.002$ adjustment. For the shots with B-field, the incident polarization was setup at $-45^{\circ}$ and $+45^{\circ}$ relative to $E_{\parallel}$ and $E_{\perp}$ respectively, yielding a ratio $R= 0.50\pm 0.01$ for the situation without B-field.

\begin{figure}
\center
\includegraphics[width=150mm]{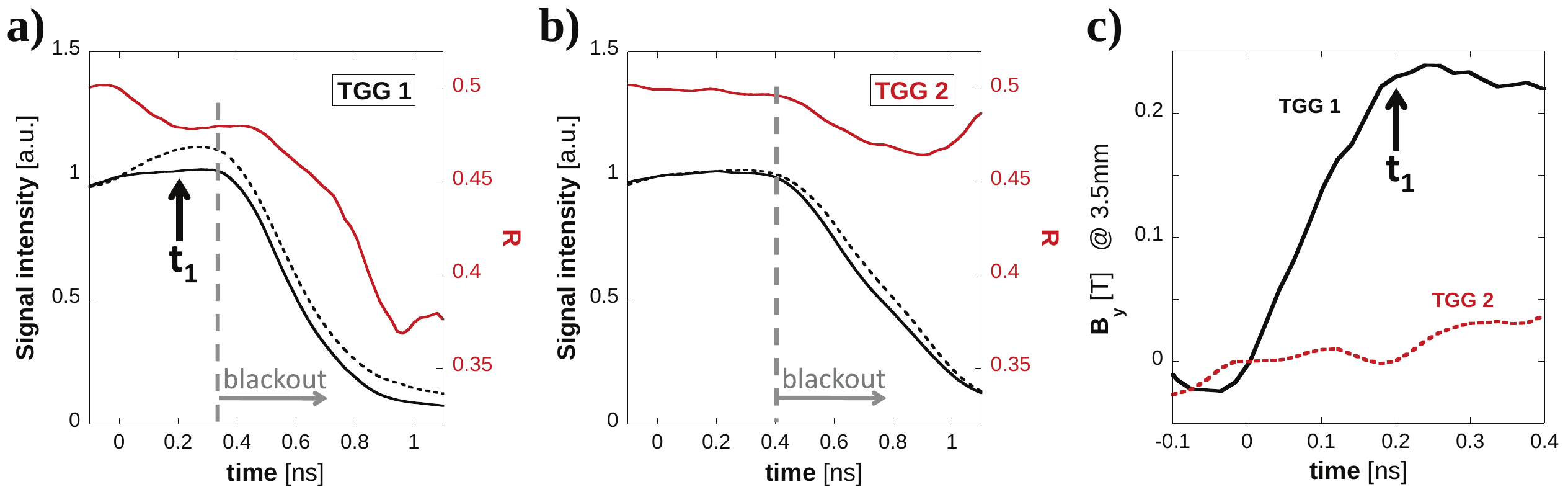}
\caption{\small (color online) Faraday rotation effect on probe laser light over two TGG crystals, placed at $3.5\mm$ from the coil plane, TGG1 centred on the coil axis and TGG2 at a $1.9\mm$ perpendicular offset, evaluated from a time-resolved polarimeter: 
{\bf a)} Detected signals of perpendicular polarizations: $E_{\parallel}^2$ (solid black) and $E_{\perp}^2$ (dashed black), and the corresponding polarization ratio $R$ (red), for TGG1.
{\bf b)} Idem, for TGG2. 
{\bf c)} Axial component of the B-field averaged over the crystals $500\mum$-thickness: solid black for TGG1, dashed red for TGG2. \label{Fig-FR}}
\end{figure}

Figure \ref{Fig-FR} shows the analysis of a successful shot for the Faraday rotation effect measurement, as a function of time: Transmitted $E_{\parallel}^2$ (solid black) and $E_{\perp}^2$ (dashed black), and corresponding $R$ (solid red) are plotted for a) TGG1 and b) TGG2 (the plotted values correspond to averages over the respective crystals' width). The crystals' blackout times are indicated by the light grey dashed vertical lines. Panel c) plots the corresponding local axial component of the B-field averaged over the crystals thickness and width, solid black for TGG1, dashed red for TGG2, obtained assuming that the TGG Verdet constant does not change with time due to the increase in crystal temperature. We see that the B-field inferred from the TGG2 measurements is too weak and remains at the signal noise level, of $\pm0.03\T$. As for the TGG1 measurements, the B-field clearly rises up to the blackout time. To compare with B-dot probe measurements we selected the time $t_1\approx 0.2\,$ns, indicated in both panels a) and c), where the averaged B-field is evaluated to $\approx 0.22 \pm 0.05\T$ along the coil axis at a $3.50\pm 0.25\,$mm distance from its centre. 
The black square symbol in Fig.\,\ref{Fig-allData}-b) corresponds to the $B_0$ strength (right-hand side ordinate axis) extrapolated from that measurement, using the same magnetostatic code\,\cite{Radia} and a similar protocol as for the B-dot data. In spite of the diagnostic uncertainties - related to the crystal thickness, accessible width and equivalent height of the streak slit on the crystal plane, and the B-field gradients over such crystal volume - the obtained value for the extrapolated $B_0$ is fairly consistent with the B-dot probe measurements.

\subsection{Proton-deflectometry}

The proton-deflectometry technique allows to measure the B-field directly in the coil region. A proton beam was created with a short laser pulse ($50\J$ on target, $1\ps$ FWHM at $1\omega_0$) focused onto $10\mum$-thick Au foils at $\approx10^{19}\Wcm$ intensity. Proton beams of $\sim 20\MeV$ maximum energy were generated by the Target Normal Sheath Acceleration (TNSA) mechanism at the foils' rear surface \cite{Snavely_PRL2000, Wilks_PoP2001}, located at a distance of $5\,$mm from the target coil. The proton beam propagation axis was perpendicular to the coil axis, and it was detected $45\,$mm away from the coil by a 15-layers radiochromic film (RCF) stack. The proton deflections due to the B-field were quantified with the help of a $42\mum$-pitch mesh, positioned $3\,$mm before the coil [see Fig.\,\ref{Fig-setup}-b)]. RCF proton imprint images correspond to a magnification of 10 for the coil plane and of 25 for the mesh plane. A $195\mum$-Al protection film before the RCF stack limited the detection to protons of energy $\epsilon_p\geq 5\MeV$ and electrons with energies $\epsilon_e\geq 260\keV$. The proton imprint signal on each RCF layer corresponds to a narrow energy range of their spectrum, due to the Bragg peak energy absorption. Accounting for the time-of-flight (TOF) between the proton source and the coil, each shot, with a chosen delay $\Delta \tau$ between the main and proton-driving lasers, scanned the effects of the B-field on the proton-trajectories over the time range of $t=\Delta \tau + \mathrm{TOF}$, with TOF between 80 and $160\,$ps.   

\begin{figure}
\center
\includegraphics[width=150mm]{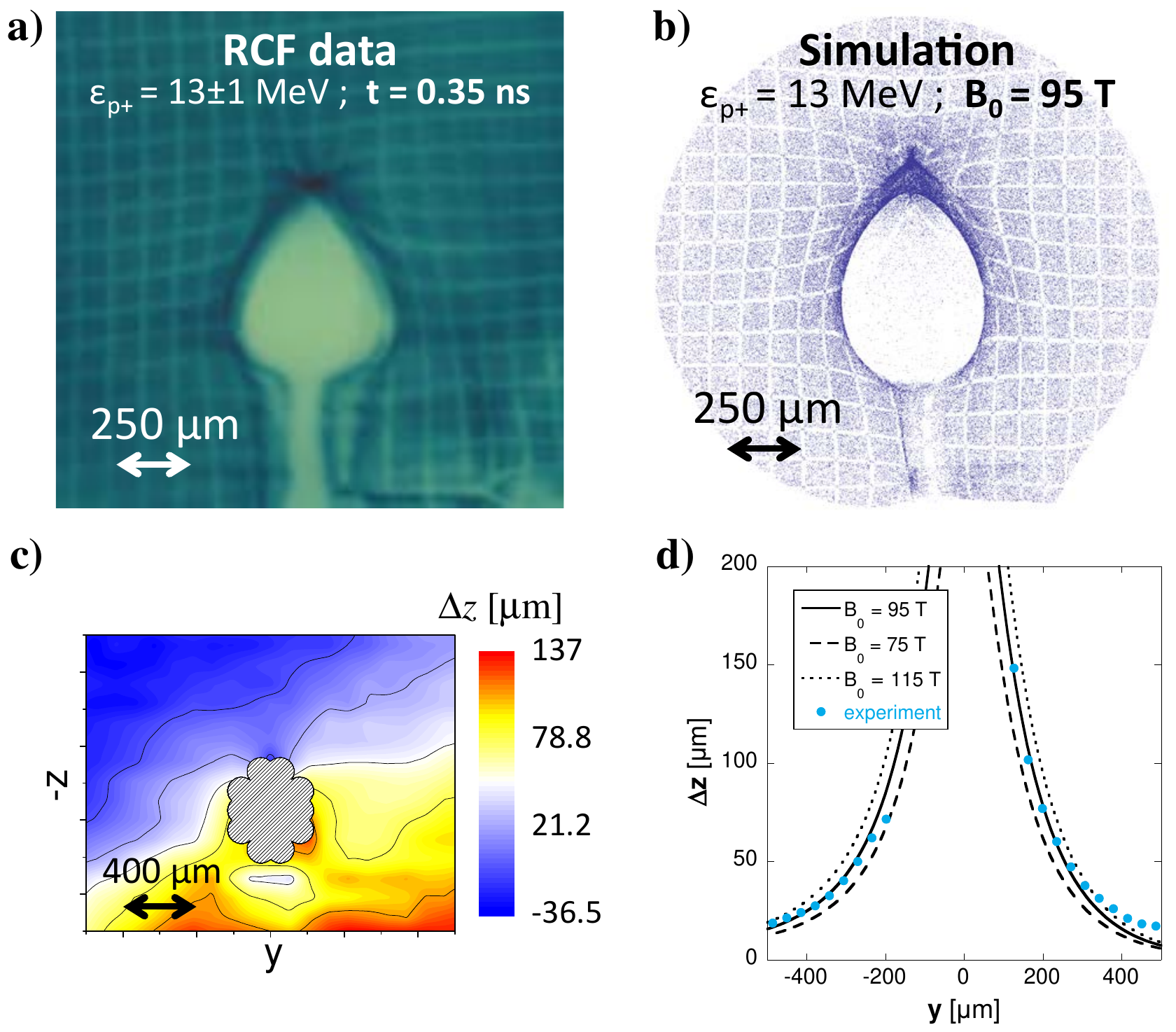}
\caption{\small (color online) {\bf a)} Sample RCF image of proton-deflectometry measurements in an early probing time $t=0.35\,$ns after ns-laser light starts to irradiate the B-loop target's rear disk. The RCF position corresponds to the imprint of $13\pm1\,$MeV protons. {\bf b)} $13\pm1\MeV$ proton imprint on detector plane given by a Monte-Carlo simulation of their trajectories over a 3D B-field map with $B_0 = 95\T$ (previously calculated by the magnetostatic code). {\bf c)} Map of experimental vertical deformations of the mesh shadow. The spatial scale on each image corresponds to the plane of the coil centre. {\bf d)} Corresponding horizontal lineouts of the experimental (circles) and synthetic (curves, for different B-field strengths) vertical deformations of the mesh imprint. \label{Fig-protons1}}
\end{figure} 

Figure \ref{Fig-protons1}-a) shows a typical image of the RCF layer corresponding to the imprint of protons of energy $\epsilon_p=13\pm1\,$MeV, obtained in a shot with $\Delta \tau = 0.25\,$ns. The corresponding probing time is $t\approx0.35\,$ns. The mesh-shadow deformations are detected for distances till $\sim 500\mum$ from the coil centre (the given spatial scale corresponds to the plane of the coil centre, perpendicular to the proton beam axis, where the unperturbed mesh shadow has a pitch of $105\mum$). However, the most outstanding feature is the centred bulb region void of any proton imprint due to the very strong B-field. Figure \ref{Fig-protons1}-b) shows the result of a Monte-Carlo simulation of the trajectories of randomly injected protons within the energy range of the experimental signal in a). Both the mesh-shadow deformations and bulb size compare very well with the experimental image when imposing a coil current $I=40\,$kA, yielding the B-field in the coil centre $B_0 = 95\,$T. Figure \ref{Fig-protons1}-c) shows the map of the experimental mesh vertical deformations [measured from the film corresponding to Fig.\,\ref{Fig-allData}-a)], fairly indicating a dipole-like B-field spatial distribution, with a typical length scale of $1\,$mm. In Fig.\,\ref{Fig-protons1}-d) the horizontal lineout of the experimental vertical deformation map at $z=0\pm 105\mum$ (circles) is compared to the corresponding synthetic deformations for different strengths of the B-field (curves), leading to the evaluation of $B_0=95\pm20\,$T at the corresponding probing time, $t\approx0.35\,$ns. This value, solid red circle in Fig.\,\ref{Fig-allData}-b) (right-hand-side ordinates), agrees with the evaluation from the B-dot measurements.

\begin{figure}
\center
\includegraphics[width=150mm]{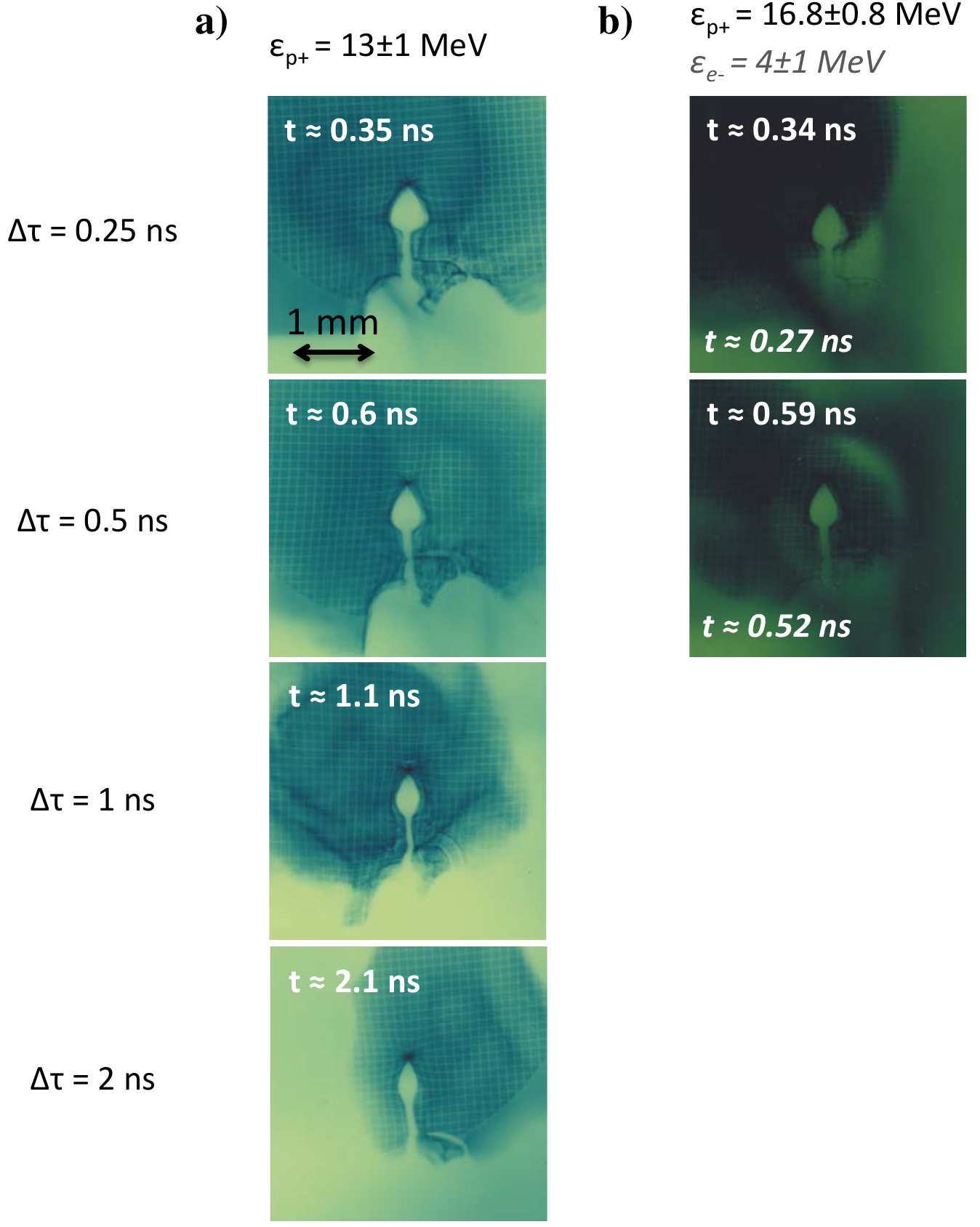}
\caption{\small (color online) Sample RCF data from different shots with varying delay $\Delta\tau$: {\bf a)} For $13\pm1\,$MeV protons, and {\bf b)} for $16.8\pm0.8\,$MeV protons and $4\pm1\,$MeV electrons. The corresponding probing times for protons are shown at the upper part of the images (straight t) and for electrons in the lower part (italic {\it t}). The spatial scale corresponds to the plane of the coil centre.  \label{Fig-protons2}}
\end{figure}

Figure \ref{Fig-protons2}-a) shows the RCF images obtained for $\epsilon_p=13\pm1\,$MeV protons with different delays $\Delta \tau$ between the laser pulses. Surprisingly, the bulb size and the overall mesh imprint deformations decrease with time: protons of the same energy undergo smaller deflections if injected at later times. The measurement of the void bulb size and mesh-imprint deformations for $t>0.35\,$ns, following the same protocol as before, would hold the $B_0$ values and uncertainties represented by the small orange circles in Fig.\,\ref{Fig-allData}-b): the $B_0$ decreasing behavior as a function of time is in contradiction with the B-dot probe measurements, and does not agree with the laser-charging process up to $t=1\,$ns. Figure \ref{Fig-protons2}-b) shows images from RCF layers corresponding to $\epsilon_p=16.8\pm0.8\,$MeV protons, for the same first two delays $\Delta\tau$: for each delay, the usual mesh imprint and void-bulb proton signatures are smaller, consistently with the higher proton energy. Yet, we detect a second particle species producing a large circular halo on both images, superposed to the proton signature. For each shot, such a halo is clearly visible with the same size and shape over the successive last six layers of the RCF stack, identifying it as the signature of relativistic electrons, accelerated at the Au-foil front surface by the short laser pulse, and relatively homogeneously deposing energy over the successive RCF layers. Accounting for their spectrum (inferred from Particle-in-Cell simulations of the short pulse laser interaction), the foil potential barrier and the Al-filtering of the RCF stack, the halo signal should correspond to electrons with energies between 3 and $5\,$MeV. Comparing images from different shots [see Fig.\,\ref{Fig-protons2}-b)], we observe that the halo size increases for increasing $\Delta\tau$. 

The signature of the relativistic electrons evolves in opposition to the proton signature. This can be explained by a monotonous increase, over the ns time scale of the main laser irradiation, of magnetized plasma electrons in regions of strong B-field around or at the vicinity of the coil. 
The main-laser interaction with the target rear disk creates a plasma $3\,$mm below the coil. As determined by X-ray spectrometry along with a comparison with calculated atomic spectra, its temperature is of the order of $T\approx 1.5\pm0.1\,$keV, with a high-energy electron component of $T_e\approx 40\pm5\,$keV. A fraction of these electrons can stream upwards to the coil-wire region in less than $30\,$ps, where they can be magnetized: Their Larmor radius is $<10\mum$ [$B_0\approx 100\T$ is achieved at $t\approx 300\,$ps (cf. Fig.\,\ref{Fig-allData})], much smaller than the typical size of the strong B-field region, $\sim 2a=500\mum$. The progressively increasing charge of magnetized electrons at the coil vicinity can produce an electrostatic effect sufficient important to influence the deflection trajectories of the probing particles, with increasing {\it focusing} or {\it defocusing} contributions respectively for the TNSA protons and for the relativistic electrons issuing from the short laser pulse interaction with the Au foil.
A trapped electron charge as small as $5\times10^{-7}\,$C yields an electrostatic potential at a distance $r\sim a$ from the coil centre already comparable to the $20\,$MeV proton maximum energy. This charge corresponds to a density of magnetized electrons $n_e^{\mathrm{mag}}\sim5\times10^{16}\cc$ in a sphere of radius $a$, which is about $5\%$ of the density of the expected supra-thermal electrons expected to stream through the coil region. This would be enough to electrostatically balance the main magnetic force on the probing TNSA protons and relativistic electrons. The magnetization condition for the main plasma electrons, $\omega_{ce}\gtrsim\omega_{pe}$ (assuring that the potential stays localized at the coil vicinity), determines a maximum $n_e^{\mathrm{mag}}\lesssim10^{17}\cc$ (for $B_0\sim100\T$), consistent with the previous estimation.

Other plasma effects can influence proton-deflectometry measurements, but their effect is expected to be negligible compared to the electrostatic one: i) The pressure of the magnetized electrons in the coil region $p_e=n_e^{\mathrm{mag}}T_e\sim 6\times10^7\,$Pa, which remains much smaller than the density of magnetic energy $p_B=B_0^2/2\mu_0\sim 4\times10^9\,$Pa, for $B_0=100\,$T. ii) Diamagnetic currents of the trapped electrons $j_1\sim en_e^{\mathrm{mag}}v_e$, yield a magnetic field over the characteristic length $a$ of $B_1\sim\mu_0j_1a\sim 10\,$T, eventually opposed but much smaller than the $B_0$ inferred by the other diagnostics. 
 
\section{Conclusions}

In conclusion, the results obtained in our experiments were obtained simultaneously by three independent diagnostics showing a reproducible quasi-static B-field generation by laser interaction with capacitor-coil targets, typically with a few ns duration and a $1\,$mm$^3$-volume, yielding peak strengths of several hundreds Tesla, depending on the target material: the dipole-like magnetic field total energy is of the order of $8.3\pm1.5\%$, $4.5\pm0.8\%$ and $0.35\pm0.05\%$ of the invested laser pulse energy, respectively for Cu, Ni and Al targets. The observed differences with the different target materials may be attributed to i) the different resistive behavior at low temperature, though the current rise time and consequent wire heating is identically rapid erasing resistivity differences, and, probably more important, to ii) the plasma temperature and hydrodynamics yielding different short-circuiting times: this will be object of further investigations.
The correct extrapolation of the B-field amplitude at the centre of the coils from distant measurements (at a few mm for Faraday rotation, and at a few cm for the pick-up coil probes) is performed by an accurate coding of the target shape and magnetostatic computation of the current intensity looping in the capacitor-coil targets. Results from Faraday rotation and proton-deflectometry are consistent with the B-dot probe measurements at the early stages of the target charging, up to $t\approx 0.35\,$ns, and then are disturbed by radiation and plasma effects, respectively the blackout of the birefringent crystal and as we concluded a negative electrostatic potential. The later is formed likely due to electron magnetization around the coil region, summing up to the main B-field effect and explaining the decreasing proton deflections for $t>0.4\,$ns.

While the typical mm$^3$-volume and ns-duration of the produced B-field pulses are small compared to the parameters achieved in the large-scale experiments and the state-of-the-art magnets mentioned in the introduction, they are characterized by an unprecedentedly high conversion efficiency (approaching the range of $10\%$) of the driver energy into magnetic energy. Moreover, this all-optical technique lends itself to the magnetization and accurate probing of high-energy-density samples driven by secondary powerful laser or particle beams: Given the $\approx 10\,\mathrm{\mu m/ns}$ expansion velocity of the coil rod, the strong B-field region is accessible for several ns.

\section*{Acknowledgments}

We gratefully acknowledge the support of the LULI pico 2000 staff during the experimental run. This work was performed through funding from the French National Agency for Research (ANR) and the competitiveness cluster Alpha - Route des Lasers, project number TERRE ANR-2011-BS04-014. The authors also acknowledge support from the COST Action MP1208 "Developing the physics and the scientific community for Inertial Fusion" through four STSM visit grants. The research was carried out within the framework of the "Investments for the future" program IdEx Bordeaux LAPHIA (ANR-10-IDEX-03-02) and of the EUROfusion Consortium, which received funding from the European Union's Horizon 2020 research and innovation program under grant agreement number 633053. The views and opinions expressed herein do not necessarily reflect those of the European Commission.
J.J.S. gratefully acknowledges fruitful discussions with L. Gremillet and K. Mima.

\section*{References}

\bibliographystyle{unsrt}

\begin{thebibliography}{}

\bibitem{Lai_RMP2001}
	D. Lai, Rev. Mod. Phys. {\bf 73}, 3, 629 -- 661 (2001). 

\bibitem{Liang_PRL2003}
	E. Liang, K. Nishimura, H. Li, and S.P. Gary, Phys. Rev. Lett. {\bf 90}, 085001 (2003).
	
\bibitem{Murdin_ncomms2013} 
	B.N. Murdin \etal, Nature Comm. 4:1469 (2013).
	
\bibitem{Lerner_JFusEn2011}
	E.J. Lerner \etal, J. Fusion Energy {\bf 30}, 5, 367 (2011).

\bibitem{Engel_PRA2008}
	D. Engel and G. Wunner, Phys. Rev. A {\bf 78}, 032515 (2008).

\bibitem{Nordhaus_PNAS2011}
	J. Nordhaus \etal, Proc. Natl Acad. Sci. USA {\bf 108}, 3135Ð3140 (2011).
	
\bibitem{Ciardi_PRL2013}
	A. Ciardi \etal, Phys. Rev. Lett. {\bf 110}, 025002 (2013).
	
\bibitem{Albertazzi_Science2014}
	B. Albertazzi \etal, Science {\bf 346}, 325 (2014).
	
\bibitem{Korneev_PoP2014}
	Ph. Korneev \etal, Phys. Plasmas {\bf 21}, 022117 (2014).
	
\bibitem{Matsumoto_Science2015}
	Y. Matsumoto, T. Amaro, T.N. Kato, M. Hoshino, Science {\bf 347}, 975 (2015).	
		
\bibitem{Perkins_PoP2013} 
	L.J. Perkins \etal, Phys. Plasmas {\bf 20}, 072708 (2013).
	
\bibitem{Strozzi_PoP2012}
	D.J. Strozzi \etal, Phys. Plasmas {\bf 19}, 072711 (2012).

\bibitem{Wang_PRL2015}
	W.-M. Wang, P. Gibbon, Z.-M. Sheng, and Y.-T. Li, Phys. Rev. Lett. {\bf 114}, 015001 (2015).
	
\bibitem{Johzaki_NuclFus2015}
	T. Johzaki \etal, Nucl. Fusion {\bf 55}, 053022 (2015).
		
\bibitem{Roth_IPAC2014}
	M. Roth \etal, Proc. of IPAC2014, Dresden, Germany.
	
\bibitem{Chen_PoP2014}
	H. Chen \etal, Phys. Plasmas {\bf 21}, 040703 (2014).

\bibitem{Debray_CRPhys2013}
	F. Debray \etal, C. R. Physique {\bf 14}, 2 (2013).
	
\bibitem{Gomez_PRL2014} 
	M.R. Gomez \etal, Phys. Rev. Lett. {\bf 113}, 155003 (2014).

\bibitem{Schmit_PRL2014}
	P.F. Schmit \etal, Phys. Rev. Lett. {\bf 113}, 155004 (2014).
	
\bibitem{Bykov_PhysB2001}	
	A.I. Bykov \etal, Physica B 294-295, 574 (2001).
	
\bibitem{Perez_PRL2013}
	F. P\'erez, A.J. Kemp, L. Divol, C.D. Chen, and P.K. Patel, Phys. Rev. Lett. {\bf 111}, 245001 (2013).
	
\bibitem{Korneev_PRE2015}
	Ph. Korneev, E. d'Humi\`eres, V. Tikhonchuk, Phys. Rev. E {\bf 91}, 043107 (2015).

\bibitem{Albertazzi_RSI2013}
	B. Albertazzi \etal., Rev. Sci. Instrum. {\bf 84}, 043505 (2013).
		
\bibitem{Daido_PRL1986}
	H. Daido \etal, Phys. Rev. Lett. {\bf 56}, 846 (1986).

\bibitem{Courtois_JAP2005} 
	C. Courtois \etal, J. Appl. Phys. {\bf 98}, 054913 (2005).
	
\bibitem{Fujioka_SciRep2013}
	S. Fujioka \etal, Scientific Reports {\bf 3}, 1170 (2013).

\bibitem{Radia}
	http://www.esrf.eu/Accelerators/Groups/InsertionDevices/Software/Radia

\bibitem{Dubois_PRE2014}
	J.-L. Dubois \etal, Phys. Rev. E {\bf 89}, 013102 (2014).
	
\bibitem{Poye_PRE2015}
	A. Poy\'e \etal, Phys. Rev. E. {\bf 91}, 043106 (2015).

\bibitem{Snavely_PRL2000}
	R.A. Snavely \etal, Phys. Rev. Lett. {\bf 85}, 14, 2945 (2000).
	
\bibitem{Wilks_PoP2001}
	S.C. Wilks \etal, Phys. Plasmas {\bf 8}, 2, 542 (2001).

\end{thebibliography}

\end{document}